\documentstyle[12pt]{article}

\catcode`\@=11
\def\section{\@startsection{section}{1}{0.0ex}
                   {3.5ex plus -1.0ex minus -0.2ex}
                   {2.3ex plus 0.2ex}{\bf}}
\def\subsection{\@startsection{subsection}{2}{0.0ex}
                        {3.25ex plus 1ex minus .2ex}
                        {1.5ex plus .2ex}{\bf}}
\catcode`@=12
\begin{document}
\def\thefootnote{\fnsymbol{footnote}}
%\FERMILABConf{94/094--T}
\begin{titlepage}
\vspace*{-0.7in}
\begin{flushright}
        FERMILAB--Conf--94/094--T\\
        OSU Preprint 288\\
        April 1994\\
\end{flushright}
\vspace{-0.40in}
\begin{center}
{\large \bf A New Bottom-Up Approach for the\\
	Construction of Fermion Mass Matrices}\footnote{Presented by C. H.
	Albright at the Yukawa Workshop held at the University of Florida,
	Gainesville, \qquad February 11 - 13, 1994.}\\
%\vfill
\vskip 0.20in
        {\bf Carl H. ALBRIGHT}\\
 Department of Physics, Northern Illinois University, DeKalb, Illinois
60115\footnote{Permanent address}\\[-0.2cm]
        and\\[-0.2cm]
 Fermi National Accelerator Laboratory, P.O. Box 500, Batavia, Illinois
60510\footnote{Electronic address: ALBRIGHT@FNALV}\\
        and\\
        {\bf Satyanarayan NANDI}\\
 Department of Physics, Oklahoma State University, Stillwater, Oklahoma
        74078\footnote{Electronic address: PHYSSNA@OSUCC}\\
\end{center}
\vskip 0.5in
%\vfill
\begin{abstract}
\indent We describe a new technique which enables one to construct an
SO(10)-symmetric fermion mass matrix model at the supersymmetric grand
unification scale directly from the fermion mass and mixing data at the low
energy scale.  Applications to two different neutrino mass and mixing scenarios
are given.
\end{abstract}
\vfill
%\noindent PACS numbers: 12.15.Ff, 12.60Jv
%\noindent $<{\rm Bulletin Board: hep-th@xxx.lanl.gov - 9403268}>$
\end{titlepage}
\section{Description of the Technique}

A new bottom-to-up approach proposed by us is summarized briefly in this talk
with more details and references presented elsewhere.\cite{AN}  We shall
restrict
our attention here to the construction of complex symmetric mass matrices
arising with Higgs in the ${\bf 10}$ and ${\bf 126}$ representations
of SO(10).  The procedure allows us to construct mass matrices which
exhibit as simple an SO(10) structure as possible with the maximum number
of texture zeros allowed for the neutrino mass and mixing scenario in question.
\begin{itemize}
\item[1)]   Start from a set of quark and lepton masses and mixing matrices
	completely specified at the low scales.
\item[2)]   Evolve the masses and mixing matrices to the GUT scale
        by making use of the one-loop renormalization group equations (RGEs)
        for the minimal supersymmetric standard model.
\end{itemize}
For this purpose we set $\Lambda_{SUSY} = 170$ GeV and $\Lambda_{GUT} = 1.2
\times 10^{16}$ GeV.  Following Naculich,\cite{Nac} we use the approximation
that only the top and bottom quarks as well as the tau lepton contribute to the
non-linear Yukawa terms in the RGEs.  With a physical top mass expected to
be near 160 GeV, we take the running mass to be $m_t(m_t) = 150$ GeV and
adjust $m_b(m_b)$ and $\tan \beta$, so consistency is achieved at
$\Lambda_{GUT}$ which requires complete Yukawa unification with
$\tan \beta \simeq 48.9$.  We are working under the assumption that only one
SO(10) $\bf{10}$ of Higgs contributes to the 33 elements of the up, down
and charged lepton mass matrices.
\begin{itemize}
\item[3)] Construct a numerical set of complex-symmetric mass matrices
 	$M^U,\ M^D,\ M^E$ and $M^{N_{eff}} = M^{N_{Dirac}}M^{-1}_R
	M^{N_{Dirac}T}$ for the up and down quarks, charged leptons and
	light neutrinos by making use of a procedure due to Kusenko,\cite{Kus}
	now applied to both quarks and leptons.
\end{itemize}
Since the quark CKM mixing matrix is unitary and represents an element of the
unitary group U(3), one can express it in terms of one Hermitian generator of
the corresponding U(3) Lie algebra times a phase parameter $\alpha$ by writing
	$$ V_{CKM} = U'_L U^{\dagger}_L = \exp(i\alpha H) \eqno(1a)$$
where
	$$i\alpha H = \sum^3_{k=1}(\log v_k){{\prod_{i\neq k}(V_{CKM}
		- v_i I)}\over{\prod_{j\neq k}(v_k - v_j)}} \eqno(1b)$$
in terms of the eigenvalues $v_j$ of $V_{CKM}$, by making use of Sylvester's
theorem.\cite{Kus}  The transformation matrices from the weak to the mass bases
are given in terms of the same generator but modified phase parameters such
that
$$U'_L = \exp(i\alpha Hx_q),\qquad U_L = \exp\left[i\alpha H(x_q - 1)\right]
	\eqno(1c)$$
and relation (1a) is preserved.
The complex symmetric quark mass matrices in the weak basis are then related
to those in the diagonal mass basis by
$$M^U = U'^{\dagger}_L D^U U'^{\dagger T}_L,\qquad M^D = U^{\dagger}_L D^D
        U^{\dagger T}_L\eqno(1d)$$
It suffices
to expand $V_{CKM},\ U'_L$ and $U_L$ to third order in $\alpha$ in order to
obtain accurate numerical results for the mass matrices $M^U$ and $M^D$.
Similar expressions can be obtained for the light neutrino and charged lepton
mass matrices from the lepton mixing matrix and its eigenvalues with $x_{\ell}$
replacing $x_q$.

The parameters $x_q$ and $x_{\ell}$ control the choice of bases for the quark
and lepton mass matrices, respectively. The up quark mass matrix is diagonal
for $x_q = 0$, while the down quark mass matrix is diagonal for $x_q = 1$;
likewise, the light neutrino mass matrix is diagonal for $x_{\ell} = 0$,
while the charged lepton mass matrix is diagonal for $x_{\ell} = 1$.

\begin{itemize}
\item[4)]   Vary $x_q$ and $x_{\ell}$ and the signs of the mass eigenvalues
	to search for simplicity in the SO(10) framework, i.e.,
	pure ${\bf 10}$ or pure ${\bf 126}$ contributions for as many mass
	matrix elements as possible.
\end{itemize}
For this purpose we note that ${\bf 10}$'s contribute equally to the down
quark and charged lepton mass matrices, while ${\bf 126}$ contributions differ
by a factor of -3; likewise for the up quark and Dirac neutrino mass matrices.
In terms of the Yukawa couplings and the appropriate VEVs, the mass matrices
are given by
$$\begin{array}{rl}
        M^U&= \sum_i f^{(10_i)}v_{ui} + \sum_j f^{(126_j)}w_{uj}\nonumber\\
        M^D&= \sum_i f^{(10_i)}v_{di} + \sum_j f^{(126_j)}w_{dj}\cr
        M^{N_{Dirac}}&= \sum_i f^{(10_i)}v_{ui} - 3\sum_j f^{(126_j)}w_{uj}\cr
        M^E&= \sum_i f^{(10_i)}v_{di} - 3\sum_j f^{(126_j)}w_{dj}\cr
                \end{array}\eqno(2)$$
\begin{itemize}
\item[5)]   For the best choice of $x_q$ and $x_{\ell}$ which maximizes the
	simplicity, construct a simple model of the mass matrices with as
	many texture zeros as possible.
\item[6)]   Evolve the mass eigenvalues and mixing matrices determined from the
        model at the SUSY GUT scale to the low scales and compare the results
        with the starting input data.
\end{itemize}

\section{Application to Two Different Neutrino Mass and Mixing Scenarios}

We now illustrate the technique by applying it to two different neutrino
scenarios, both of which explain the solar neutrino depletion data with
the non-adiabatic Mikheyev-Smirnov-Wolfenstein\cite{MSW} (MSW) effect, where
one includes the atmospheric neutrino depletion phenomenon\cite{atm} while the
other accepts the cocktail model\cite{cock} interpretation of missing dark
matter.

We take the same quark input data for both models. For the light quark
masses, we shall adopt the values quoted by Gasser and Leutwyler\cite{GL} while
the heavy quark masses are specified at their running mass scales:
$$\begin{array}{rlrl}
        m_u(1 {\rm GeV})&= 5.1\ {\rm MeV},& \qquad m_d(1 {\rm GeV})&= 8.9\
                {\rm MeV}\nonumber\\
        m_c(m_c)&= 1.27\ {\rm GeV},& \qquad m_s(1 {\rm GeV})&= 175\ {\rm MeV}
                \cr
        m_t(m_t)&= 150\ {\rm GeV},& \qquad m_b(m_b)&\simeq 4.25\ {\rm GeV}\cr
  \end{array}\eqno(3a)$$
For the Cabibbo-Kobayashi-Maskawa (CKM) mixing matrix, we adopt the following
central values at the weak scale
$$V_{CKM} = \left(\matrix{0.9753 & 0.2210 & (-0.283 -0.126i)\times 10^{-2}\cr
                -0.2206 & 0.9744 & 0.0430\cr
                0.0112 -0.0012i & -0.0412 -0.0003i & 0.9991\cr}\right)
        \eqno(3b)$$
where we have assumed a value of 0.043 for $V_{cb}$ and applied strict
unitarity to determine $V_{ub},\ V_{td}$ and $V_{ts}$.

\subsection{Lepton Scenario (A) involving the Non-adiabatic MSW Solar and
	Atmospheric Neutrino Depletion Effects}

In this scenario, we single out the central points in the two neutrino mixing
planes
$$\begin{array}{rlrl}
	\delta m^2_{12}&= 5 \times 10^{-6}\ {\rm eV}^2,&\qquad
		\sin^2 2\theta_{12}&= 8 \times 10^{-3}\nonumber\\
	\delta m^2_{23}&= 2 \times 10^{-2}\ {\rm eV}^2,&\qquad
		\sin^2 2\theta_{23}&= 0.5\cr \end{array}\eqno(4)$$
With the neutrino masses assumed to be non-degenerate, we
take for the lepton input
$$\begin{array}{rlrl}
        m_{\nu_e}&= 0.5 \times 10^{-6}\ {\rm eV},& \qquad m_e&= 0.511\ {\rm
                MeV}\nonumber\\
        m_{\nu_{\mu}}&= 0.224 \times 10^{-2}\ {\rm eV},& \qquad m_{\mu}&=
                105.3\ {\rm MeV}\cr
        m_{\nu_{\tau}}&= 0.141\ {\rm eV},& \qquad m_{\tau}&= 1.777\ {\rm
                GeV}\cr \end{array}\eqno(5a)$$
and
$$V^{(A)}_{LEPT} = \left(\matrix{0.9990 & 0.0447 & (-0.690 -0.310i)
		\times 10^{-2}\cr -0.0381 -0.0010i & 0.9233 & 0.3821\cr
                0.0223 -0.0030i & -0.3814 & 0.9241\cr}\right) \eqno(5b)$$
where we have simply assumed a value for the electron-neutrino mass to
which our analysis is not very sensitive and constructed
the lepton mixing matrix by making use of the unitarity conditions with
the same phase in (5b) as in (3b).

We evolve the masses and mixing matrices to the SUSY GUT scale and
use the extended Kusenko\cite{Kus} procedure to construct the mass matrices
numerically. The simplest SO(10) structure for the mass matrices is found
with $x_q = 0$ and $x_{\ell} = 0.88$, in which case the matrices have the
following SO(10) transformation properties:
$$M^U \sim M^{N_{Dirac}} \sim diag(126;\ 126;\ 10)\eqno(6a)$$
$$M^D \sim M^E \sim \left(\matrix{10',126 & 10',126' & 10'\cr 10',126' &
	126 & 10'\cr 10' & 10' & 10\cr}\right)\eqno(6b)$$
Note that the same ${\bf 10}$ is assumed to contribute to the 33 elements
of the above mass matrices, with Yukawa coupling unification achieved for
$\tan \beta \simeq 48.9$.

In this scenario we are able to construct a simple SO(10) model with just
nine independent parameters for the following four matrices, such that
$$\begin{array}{rlrl}
        M^U&= diag(F',\ E',\ C')& \qquad M^{N_{Dirac}}&= diag(-3F',\ -3E',\ C')
		\nonumber\\[0.1cm]
        M^D&= \left(\matrix{0 & A & D\cr A & E & B\cr D & B & C\cr}\right)
		& \qquad
        M^E&= \left(\matrix{F & 0 & D\cr 0 & -3E & B\cr D & B & C\cr}\right)
		\end{array}\eqno(7a)$$
where only $D$ is complex and
$$C'/C = v_u/v_d,\qquad E'/E = -4F'/F = w_u/w_d \eqno(7b)$$
in terms of the ratios of the ${\bf 10}$ and of the ${\bf 126}$ vacuum
expectation values associated with the diagonal Yukawa couplings.  In this
model for
scenario (A), two ${\bf 10}$'s and two ${\bf 126}$'s are required as indicated
in (6a,b), while four texture zeros appear in the two matrices for $M^U$ and
$M^D$ and for $M^{N_{Dirac}}$ and $M^E$.

With
$$F' = -1.098 \times 10^{-3},\qquad E' = 0.314,\qquad C' = 120.3 \eqno(8a)$$
$$\begin{array}{rlrl}
        C&= 2.4607,& \qquad &{\rm so}\quad v_u/v_d = \tan \beta = 48.9
                \nonumber\\
        E&= -0.3830 \times 10^{-1},& \qquad &{\rm hence}\quad w_u/w_d =
-8.20\cr
        F& = -0.5357 \times 10^{-3},& \qquad B&= 0.8500 \times 10^{-1}\cr
        A& = -0.9700 \times 10^{-2},& \qquad D&= (0.4200 + 0.4285i) \times
                10^{-2}\cr \end{array}\eqno(8b)$$
in GeV, the masses and mixing matrices are calculated at the GUT scale by use
of the projection operator technique of Jarlskog\cite{Jarl} and then evolved to
the low scales.  The following low-scale results emerge for the quarks:
$$\begin{array}{rlrl}
        m_u(1 {\rm GeV})&= 5.10\ {\rm MeV},& \qquad m_d(1 {\rm GeV})&= 9.33
                \ {\rm MeV}\nonumber\\
        m_c(m_c)&= 1.27\ {\rm GeV},& \qquad m_s(1 {\rm GeV})&= 181\ {\rm MeV}
                \cr
        m_t(m_t)&= 150\ {\rm GeV},& \qquad m_b(m_b)&= 4.09\ {\rm GeV}\cr
  \end{array}\eqno(9a)$$
$$V_{CKM} = \left(\matrix{0.9753 & 0.2210 & (0.2089 -0.2242i)\times 10^{-2}\cr
                -0.2209 & 0.9747 & 0.0444\cr
                0.0078 -0.0022i & -0.0438 -0.0005i & 0.9994\cr}\right)
        \eqno(9b)$$
which are to be compared with the input starting data given in (3a) and (3b).

In the absence of any VEVs coupling the left-handed neutrino
fields together, we observe that the heavy righthanded Majorana neutrino mass
matrix can be computed at the GUT scale from the seesaw mass
formula
$$M^R = - M^{N_{Dirac}}(M^{N_{eff}})^{-1}M^{N_{Dirac}}\eqno(10a)$$
which can be well approximated by the nearly geometric form
$$M^R = \left(\matrix{F'' & - {2\over{3}}\sqrt{F''E''} &
                -{1\over{3}}\sqrt{F''C''}e^{i\phi_{D''}}\cr
                - {2\over{3}}\sqrt{F''E''} & E'' &
                        -{2\over{3}}\sqrt{E''C''}e^{i\phi_{B''}}\cr
                -{1\over{3}}\sqrt{F''C''}e^{i\phi_{D''}} &
                -{2\over{3}}\sqrt{E''C''}e^{i\phi_{B''}} & C''\cr}\right)
                        \eqno(10b)$$
where $E'' = {2\over{3}}\sqrt{F''C''}$ and $\phi_{B''} = - \phi_{D''}/3$.
In terms of the three additional parameters $C'' = 0.6077 \times 10^{15},
\ F'' = 0.1745\times 10^{10}$ and
$\phi_{D''} = 45^o$, the resulting heavy Majorana neutrino masses are
determined to be
$$\begin{array}{rl}
M_{R_1}&= 0.249 \times 10^9\ {\rm GeV} \nonumber\\
M_{R_2}&= 0.451\times 10^{12}\ {\rm GeV} \cr
M_{R_3}&= 0.608\times 10^{15}\ {\rm GeV} \cr \end{array}\eqno(10c)$$

{}From the model parameters, the seesaw formula and Jarlskog's projection
operator technique,\cite{Jarl} the light lepton masses and their mixing matrix
can be constructed at the GUT scale and then evolved downward to the low
scales where we find
$$\begin{array}{rlrl}
        m_{\nu_e}&= 0.534 \times 10^{-5}\ {\rm eV},& \qquad m_e&= 0.504\ {\rm
                MeV}\nonumber\\
        m_{\nu_{\mu}}&= 0.181 \times 10^{-2}\ {\rm eV},& \qquad m_{\mu}&=
                105.2\ {\rm MeV}\cr
        m_{\nu_{\tau}}&= 0.135\ {\rm eV},& \qquad m_{\tau}&= 1.777\ {\rm
                GeV}\cr \end{array}\eqno(11a)$$
and
$$V_{LEPT} = \left(\matrix{0.9990 & 0.0451 & (-0.029 -0.227i) \times 10^{-2}
		\cr -0.0422 & 0.9361 & 0.3803\cr
                0.0174 -0.0024i & -0.3799 -0.0001i & 0.9371\cr}\right)
                \eqno(11b)$$
The agreement with the initial input values in (5a,b) is excellent.

\subsection{Lepton Scenario (B) involving the Non-adiabatic MSW Solar Depletion
	Effect and the Cocktail Model for Mixed Dark Matter}

In this scenario, the parameters of the 23 mixing plane given in (4) are
replaced by
$$\delta m^2_{23} = 49\ {\rm eV}^2, \qquad
		\sin^2 2\theta_{23} = 10^{-3}\eqno(12a)$$
with the tau-neutrino now assumed to account for the 30\% hot dark matter
component of the cocktail model\cite{cock} for mixed dark matter with a mass of
$m_{\nu_{\tau}} = 7.0$ eV.  The mixing angle has been set close to the
present upper bound from accelerator experiments, so the mixing matrix is
now given by
$$V^{(B)}_{LEPT} = \left(\matrix{0.9990 & 0.0447 & (-0.289 -0.129i)
		\times 10^{-2}\cr -0.0446 & 0.9989 & 0.0158\cr
                0.0036 -0.0013i & -0.0157 & 0.9998\cr}\right) \eqno(12b)$$

Following the same general procedure as applied for scenario (A), we find the
simplest SO(10) structure is obtained for $x_q = 0.5$ and $x_{\ell} = 0.0$,
but only one texture zero is then present.  Instead we make use of the
bases where $x_q = 0$ and $x_{\ell} = 0.3$ as in scenario (A).  The quark
mass matrices are then exactly the same as before, while the SO(10) structures
are more complicated as seen from
$$M^U \sim M^{N_{Dirac}} \sim diag(10,126;\ 126;\ 10)\eqno(13a)$$
$$M^D \sim M^E \sim \left(\matrix{10,10',126 & 10',126' & 10'\cr 10',126' &
	126 & 10',126'\cr 10' & 10',126' & 10\cr}\right)\eqno(13b)$$

Without repeating all the details, we simply write down the form of the
model matrices derived in this scenario and find
$$\begin{array}{rlrl}
        M^U&= diag(F',\ E',\ C')& \qquad M^{N_{Dirac}}&= diag(-2.5F',\ -3E',
		\ C')
		\nonumber\\[0.1cm]
        M^D&= \left(\matrix{0 & A & D\cr A & E & B\cr D & B & C\cr}\right)
		& \qquad
        M^E&= \left(\matrix{{5\over{6}}F & -{1\over{3}}A & D\cr -{1\over{3}}A
		& -3E & {1\over{3}}B\cr D & {1\over{3}}B & C\cr}\right)
		\end{array}\eqno(14)$$
where again only $D$ is complex.
In this model for scenario (B), two ${\bf 10}$'s and two ${\bf 126}$'s are
required as indicated
in (13a,b), while four texture zeros appear in the two matrices for $M^U$ and
$M^D$ but only three texture zeros appear for $M^{N_{Dirac}}$ and $M^E$.
Since the quark mass matrices must lead to the same numerical
results as in scenario (A), the values for the parameters introduced above are
those given in (8a) and (8b).

The heavy right-handed Majorana mass matrix can be approximated
by the nearly geometric form
$$M^R = \left(\matrix{F'' & - 2 \sqrt{F''E''} &
                - \sqrt{F''C''}e^{i\phi_{D''}}\cr
                - 2 \sqrt{F''E''} & - E'' &
                        2 \sqrt{E''C''}e^{i\phi_{B''}}\cr
                - \sqrt{F''C''}e^{i\phi_{D''}} &
                2 \sqrt{E''C''}e^{i\phi_{B''}} & C''\cr}\right)
                        \eqno(15a)$$
where $E'' = {1\over{8}}\sqrt{F''C''}$, aside from an overall sign.
With $C'' = 0.2323 \times 10^{14},\ F'' = 0.1096\times 10^{11}$ and
$\phi_{D''} = 18.7^o$, $\phi_{B''} = 23.0^o$ and $\phi_{C''} = 41.8^o$, we find
the resulting heavy Majorana neutrino masses for this case are determined to be
$$\begin{array}{rl}
M_{R_1}&= 0.841 \times 10^9\ {\rm GeV} \nonumber\\
M_{R_2}&= 0.312\times 10^{12}\ {\rm GeV} \cr
M_{R_3}&= 0.235\times 10^{14}\ {\rm GeV} \cr \end{array}\eqno(15b)$$

By again making use of the simplified
matrices at the GUT scale first to compute the lepton masses and mixing matrix
$V_{LEPT}$ by the projection operator technique of Jarlskog and then
to evolve the results to the low scales, we find at the low scales for the
(B) scenario
$$\begin{array}{rlrl}
        m_{\nu_e}&= 0.544 \times 10^{-6}\ {\rm eV},& \qquad m_e&= 0.511\ {\rm
                MeV}\nonumber\\
        m_{\nu_{\mu}}&= 0.242 \times 10^{-2}\ {\rm eV},& \qquad m_{\mu}&=
                107.9\ {\rm MeV}\cr
        m_{\nu_{\tau}}&= 6.99\ {\rm eV},& \qquad m_{\tau}&= 1.776\ {\rm
                GeV}\cr \end{array}\eqno(16a)$$
and
$$V_{LEPT} = \left(\matrix{0.9992 & 0.0410 & (0.150 -0.107i) \times 10^{-2}
		\cr -0.0411 & 0.9991 & 0.0113\cr
                -0.0010 -0.0011i & -0.0123 & 0.9999\cr}\right)
                \eqno(16b)$$
to be compared with the initial low scale input in (5a) and (12).

\section{Summary}

In this talk we have sketched a procedure which enables one to construct
fermion mass matrices at the GUT scale which yield the low energy data taken
as input.  The models constructed for the two neutrino scenarios work well
in the SO(10) SUSY GUT framework with relatively few parameters, but the
structure exhibited for scenario (B) with a 7 eV tau-neutrino is not as
simple as that for scenario (A) based on the observed muon-neutrino
atmospheric depletion effect.  Discrete family symmetries giving rise to
these models are now under investigation.  Our general procedure to
construct mass matrices can be applied to other symmetry-based frameworks
as well.\\[0.1in]

The research of CHA was supported in part by Grant No. PHY-9207696 from the
National Science Foundation, while that of SN was supported in part by the
U.S. Department of Energy, Grant No. DE-FG05-85ER 40215.

\end{document}